# Burn-After-Use for Preventing Data Leakage through a Secure Multi-Tenant Architecture in Enterprise LLM

Qiang Zhang
School of Informatics, Computing & Cyber Systems, Steve Sanghi College of Engineering, Northern Arizona University, Flagstaff, AZ, USA
qz86@nau.edu

Elena Emma Wang
American Heritage Academy
elena.wang.mathlete@gmail.com

Jiaming Li
School of Informatics, Computing & Cyber Systems, Steve Sanghi College of Engineering, Northern Arizona University, Flagstaff, AZ, USA
jiaming.li@nau.edu

Xichun Wang
School of Informatics, Computing & Cyber Systems, Steve Sanghi College of Engineering, Northern Arizona University, Flagstaff, AZ, USA
xw247@nau.edu

## ABSTRACT

This study presents a Secure Multi-Tenant Architecture (SMTA) combined with a novel concept Burn-After-Use (BAU) mechanism for enterprise LLM environments that can effectively prevent data leakage. As institutions increasingly adopt large language models (LLMs) across departments, the risks of data leakage have become a critical security and compliance concern. Conventional institutional safeguards primarily rely on access control and encryption to mitigate unauthorized access, but they fail to address contextual persistence in the retention of sensitive information in conversational memory. The proposed SMTA isolates LLM instances across departments and enforces rigorous context ownership boundaries within an internally deployed infrastructure. The BAU mechanism introduces data confidentiality by enforcing ephemeral conversational contexts that are automatically destroyed after use, preventing cross-session or cross-user inference. The evaluation to SMTA and BAU is through two sets of realistic and reproducible experiments comprising of 127 test iterations. One aspect of this experiment is to assess prompt-based and semantic leakage attacks in a multi-tenant architecture (Appendix A) across 55 infrastructure-level attack tests, including vector-database credential compromise and shared logging pipeline exposure. SMTA achieves 92% defense success rate, demonstrating strong semantic isolation while highlighting residual risks from credential misconfiguration and observability pipelines. Another aspect is to evaluate the robustness of BAU under realistic failure scenarios (Appendix B) using four empirical metrics: Local Residual Persistence Rate (LRPR), Remote Residual Persistence Rate (RRPR), Image Frame Exposure Rate (IFER), and Burn Timer Persistence Rate (BTPR). Across 72 test iterations, BAU achieves an average success rate of 76.75%, providing strong protection against post-session leakage across client-side, server-side, application-level, infrastructure-level, and cache-level threat surfaces. These empirical results show that SMTA and BAU together enforce strict isolation and complete session ephemerality and a highly secure foundation for enterprise LLM deployments, delivering quantifiable guarantees of confidentiality, non-persistence, and policy-aligned behavior.

## 1. Introduction

Large language models (LLMs) have demonstrated strong performance across a wide range of tasks and have highly improved the work efficiency and capabilities among institutions, enterprises, and governmental administrations [1]. However, with the widespread adoption of LLMs, new security and privacy risks have emerged. In shared LLM inference environments, reasoning outputs produced during cross-departmental interactions may give rise to unintended data leakage. Recent work shows that LLMs may inadvertently memorize sensitive information, leak user data or allow unintended context propagation across sessions [2,3]. Conventional enterprises have been safeguarding by access control lists (ACLs), encryption, and network segmentation, primarily focusing on restricting external threats. However, they do not adequately mitigate internal risks arising from contextual persistence, prompt-based inference, cross-session contamination, or retrieval boundary violations [4,5]. As a result, confidential information shared from one department's LLM session may inadvertently influence responses generated for another department. This causes the data-security threats and compliance concerns. The modern AI governance frameworks such as the NIST AI Risk Management Framework [6] provides systematic approaches and structured methodology to manage such risks.

Through the analysis of recent multi-tenant LLM serving systems, Yang et al. [7] focus on improving KV-cache reuse across similar requests. They introduce a dual-stage deviation algorithm and a cache-aware scheduler that significantly reduces TTFT while maintaining model accuracy. Hu et al. [8] observe that many fine-tuned LLMs share large portions of the same foundation model and argue that treating each model as a monolithic unit is inefficient in terms of memory, storage, and computation. Their system, BlockLLM, partitions models into reusable blocks, achieving higher throughput and lower latency in multi-tenant cloud environments. Shen et al. [9] study multi-tenant LLM serving on resource-limited edge devices using parameter-efficient LoRA adapters. They introduce EdgeLoRA, which performs adaptive adapter selection and heterogeneous memory management to support thousands of adapters with improved throughput and low latency. These studies together highlight three complementary directions in multi-tenant LLM optimization: token-level sharing (KV cache), model-component sharing (blocks), and adapter-level sharing (LoRA). They improve performance but still assume shared model pipelines or shared KV-cache layers, which remain vulnerable to cross-tenant influence at the semantic level. In contrast, our Secure Multi-Tenant Architecture (SMTA) applies strict isolation across model instances, vector stores, and policy paths, preventing both semantic and retrieval-level leakage.

Permissioned LLMs allow models to enforce user-specific permissions during content generation through parameter-efficient fine-tuning techniques. Jayaraman et al. [10] propose adapter-based mechanisms: Activate, Merge, and Union that regulate model behavior according to the user's domain privileges, by modifying transformer layers with domain-specific knowledge. The approach defines formal metrics to measure how well access control is enforced but lacks structural separation between organizational tenants. Mandalawi et al. [11] automate the governance of data access requests in an enterprise context by using the policy-aware strategy based on six-stage reasoning pipeline combined with hard policy gates. Their system ensures a safe and auditable cross-department by outputting one of the three decisions: Approve, Deny, and Conditional for data-access flow. Zeng et al. [12] propose a privacy-preserving strategy, PrivacyRestore, that first removes sensitive spans from the input before inference. They then construct a weighted meta restoration vector with added noise, which, together with the sanitized input, is processed by the LLM. Finally, the model restores the missing information by steering the activations using this vector.

Zhou et al. [13] operationalize data minimization in LLM prompting by proposing practical strategies for reducing unnecessary personal information in user inputs. Their work quantifies the trade-offs between privacy and response utility, and explores prompt transformations and filtering techniques that minimize exposure of sensitive data while maintaining useful model outputs. Wang et al. [14] design an attack method called Memory EXTRaction to reveal how private information in LLM agent memory can be leaked. From the vulnerabilities exposed, they propose three categories of safeguards: designer-side, workflow-level, and access-level, which strengthen privacy by limiting memory storage and retrieval and enforcing stricter agent behaviors that reduce unintended disclosure. Kan et al. [15] safeguard sensitive user information from cloud-based conversational LLMs by employing text sanitization around pre-conversing and post-conversing. Their approach categorizes sensitive entities and applies a configurable sanitization policy that masks or abstracts them, retaining sufficient information for the conversational model to function while reducing privacy risk.

Therefore, the study to this security-oriented LLM deployment architecture combines two key mechanisms: Secure Multi-Tenant Architecture (SMTA) and Burn-After-Use (BAU) session semantics. SMTA isolates LLM model instances, retrieval storage, and policy enforcement units across tenants, building a robust security environment [16]. BAU further enforces data confidentiality by ensuring that conversational context and temporary embeddings are ephemeral and automatically destroyed after use, preventing unintended long-term retention or cross-user inference.

The assessment of this architecture is conducted through a set of behavioral experiments that target the most common and high-risk leakage vectors in enterprise LLM systems, including cross-tenant leakage attacks and burn-after-use session persistence tests. By the use of gpt-oss-120b, DeepSeek-V3.2-Exp and Mistral-7B-Instruct-v0.2 to perform these experiments [17,18,19], the results finally demonstrated the SMTA and the BAU are highly effective for the practicality and robustness of the design of this paper.

## 2. Methodologies

### 2.1 Secure Multi-Tenant Architecture (SMTA)

#### 2.1.1. LLM Private Deployment and Multi-Tenant Access Control

The Secure Multi-Tenant Architecture (SMTA) is designed to operate within a privately deployed large language model (LLM) environment, fully isolated from external public networks and third-party APIs. By hosting the model within institutional boundaries, either on-premise or in a segmented private cloud, the enterprise retains full control over the LLM's data flow, model updates, and access governance. This deployment strategy eliminates external data dependencies and prevents inadvertent exposure of confidential information to the public.

To enforce confidentiality within this private domain, SMTA employs a multi-tenant access-control framework that maps organizational departments (for example, Human Resources, Finance, or Research Development) to independent logical tenants. Each tenant operates under a least-privilege policy, ensuring that users and processes can access only those datasets and contextual histories explicitly authorized for their department. The framework integrates both Role-Based Access Control (RBAC) and Attribute-Based Access Control (ABAC) mechanisms to provide fine-grained authorization rules capable of reflecting institutional hierarchies and data-classification levels. Figure 1 illustrates the architecture workflow and the stream path for the multi-tenant LLMs.

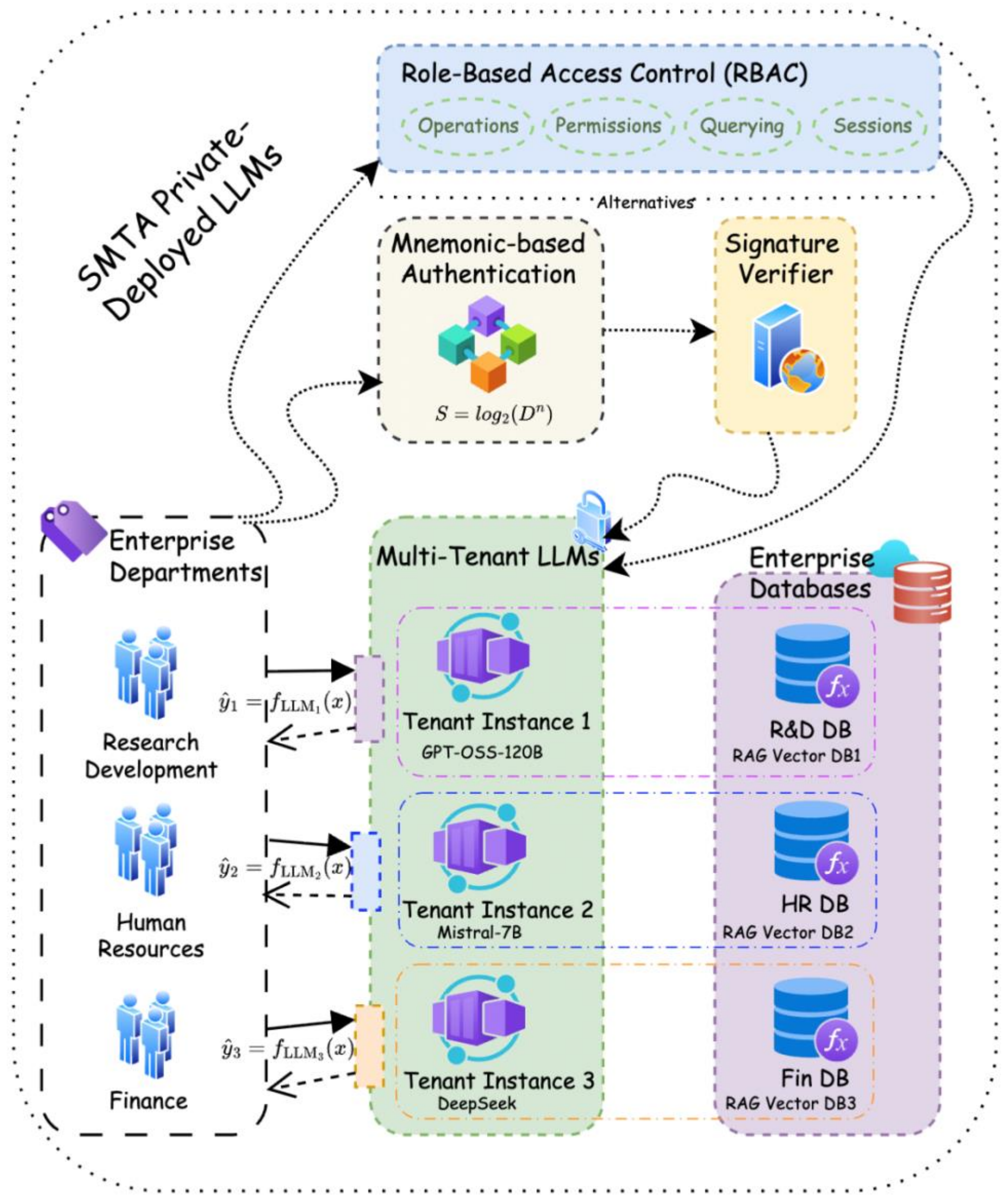

Figure 1. The Secure Multi-Tenant Architecture (SMTA) in a private-deployed LLM environment. The design of the authentication via RBAC or mnemonic-based, and the LLM instance isolation to ensure strict data governance and confidentiality. Each department operates within its own logical tenant.

Each LLM tenant instance, whether a separate container, virtual machine, or private cloud, interacts with a dedicated memory space and transient storage context. No conversational histories or vector embeddings are shared across tenants, effectively preventing cross-department inference. Authentication tokens and session identifiers are issued per tenant and expire after predefined intervals, aligning with the Burn-After-Use principle introduced in section 2.2.2. This ensures that both model contexts and user sessions remain ephemeral and non-persistent beyond their intended operational window.

### 2.1.2. Instances Separation in Inter-Departments

A critical aspect of deploying Large Language Models (LLMs) in an enterprise or institutional environments is understanding the data lifecycle under different modes of access. Public conversational interfaces such as ChatGPT, Mistral AI and DeepSeek, are designed for end-user interaction and typically retain user content for model improvement, personalization, and service analytics. The persistence of raw or partially processed data presents a non-trivial risk of contextual data leakage. API service providers, including OpenAI, DeepSeek and Mistral, have stated in their documentation that input and output data transmitted via API endpoints are not retained for long-term model training. Instead, such data may be temporarily cached for limited operational purposes, such as Service integrity and debugging, short-term quality assurance, and request deduplication or caching. However, despite these policy assurances, it is important to note that "ephemeral caching" does not equate to full data non-persistence. Cached inference results may still be temporarily stored in provider-controlled environments when invoking a public LLM through an external API, thereby potentially posing data leakage risks.

Therefore, the SMTA Architecture proposed, as Figure 1 shows above, eliminates this risk by deploying LLM instances in self-hosted, non-networked public environments, ensuring that conversational data isolation is never exposed publicly. While physically distinct instances can provide additional compartmentalization, this approach comes with significant computational and financial overhead. Large models such as *gpt-oss-120b*, *deepseek-v3.2-Exp* or *mistral-7B-instruct-v0.2* require high-end GPUs, along with a complete server infrastructure, which can cost tens to hundreds of thousands of USD per instance. The deployment of the large models is by Cloudflare, AWS or Google Cloud for launching out-of-the-box, which can be accomplished through containerized runtimes, memory segmentation, and ephemeral session management. It is feasible for multi-instance deployment and for departmental separation. The logical isolation is compulsory for a single instance or multiple instances. Each department's conversational context, embeddings, and session data are strictly separated. Table 1 shows the contrast of different modes of LLMs below.

| Access Mode | Data Retention Policy | Caching Type | Risk of Leakage | Deployment Cost | LLM Providers |
|---|---|---|---|---|---|
| Public Conversational Interface | Long-term (used for model training and analytics) | Persistent | High | Low (cloud-based) | ChatGPT, DeepSeek, Mistral (Le Chat) |
| API Invocation | Short-term (temporary cache for QA/debugging) | Ephemeral | Moderate | Low to medium | OpenAI API, DeepSeek, Mistral API |
| Self-Hosted SMTA | No external retention | Localized runtime cache temporarily | Minimal, nearly none | Medium | Enterprise LLM deployment |

Table 1: The comparison among multiple LLM providers, the costs for different deployments, the probability of data leakage while using LLMs in the business and across departments. The self-hosted LLM through SMTA is superior in privacy.

### 2.1.3 Mnemonic-Based Authentication Replacing Conventional Account-Password Pairs

Conventional authentication systems in enterprise environments typically rely on static username–password pairs stored in centralized databases. Although such methods have been widely adopted, they inherently introduce vulnerabilities associated with credential theft, replay attacks, and compromised identity stores. In a Secure Multi-Tenant Architecture (SMTA), where strict compartmentalization between tenants is critical, centralized credential repositories represent potential single points of failure. To address these issues, a mnemonic-based decentralized authentication mechanism is proposed, inspired by cryptographic practices in decentralized systems such as Web3. Instead of persistent account credentials, users are assigned a 12-word mnemonic phrase, which encodes a 128-bit entropy seed. The mnemonic serves as the sole authentication material, locally derived on the client side and never transmitted or stored on any centralized server. Below shows that S denotes the entropy strength of a mnemonic phrase composed of n words selected from a standardized dictionary of D entries. The total entropy can be expressed as:

$$S = log_2 \left(D^n\right)$$

Using the standard BIP-39 dictionary of 2048 words, a 12-word phrase yields an entropy strength of:

$$S = log_2\left(2048^{12}\right) = 12 * 11 = 132 \text{ bits}$$

This entropy level is comparable to the security of a 256-bit symmetric encryption key, sufficient to resist brute-force attacks even under advanced computational or quantum scenarios. Therefore, the mnemonic-based authentication aligns with the principles of least data retention and tenant isolation central to the SMTA. It removes the dependency on central identity services and mitigates risks of cross-tenant identity leakage.

### 2.2 Burn-After-Use Mechanism

Conventional enterprise security controls, such as role-based access control (RBAC), encrypted communication channels, and network segmentation, reduce unauthorized data exfiltration but do not eliminate context persistence due to databases. When an LLM processes sensitive enterprise documents, the extracted content is retained as conversational memory within the application layer or the model serving infrastructure, creating a latent leakage channel even when explicit data access restrictions remain intact. The Burn-After-Use mechanism introduces a non-persistent operational context in which any document-derived or user-contributed content is automatically destroyed after conversation completion, timeout, or intentional user-triggered termination. Therefore, it prevents post-conversation inference attacks, prompt extraction attacks, and session replay in model history. The mechanism is incorporated into the Secure Multi-Tenant Architecture (SMTA) as a session-layer guarantee. When a document (e.g., PDF, XLS, TXT, DOCX) is introduced into a private or public LLM environment, its contents are first parsed into plain text by an authorized preprocessing pipeline. The model itself never receives the raw document file but only the minimal tokenized representation required for the task. This parsed representation is then stored in a short-lived temporary context memory environment that is cryptographically bound to the session identity and time window. It therefore ensures that sensitive documents are "used but not kept", aligning with data minimization principles, data-handling accountability, and life-cycle governance.

#### 2.2.1 Preset Burning Intervals in Public LLM Systems

When interacting with public LLM systems such as ChatGPT, DeepSeek or Mistral AI, users cannot directly control server-side retention policies. However, confidentiality can still be partially guaranteed through client-side enforceable temporal keying. We can employ time-based key derivation, where decryption keys for document content are cryptographically tied to a specific temporal validity period. The document is symmetrically encrypted prior to upload or processing, and the decryption key is generated using a time-dependent function, such as:

$$K_t = H(seed \,||\, T_{valid} \,||\, contextID)$$

where seed is a private entropy source, $T_{valid}$ denotes the authorized temporal access window (e.g., 30 minutes, 1 hour, or 24 hours), and $contextID$ is a session identifier. When $T_{valid}$ expires, the key can no longer be recomputed, thereby rendering the encrypted content irretrievable even if the ciphertext remains cached or stored server-side. This strategy does not rely on the LLM platform offering any secure deletion guarantees. Instead, it ensures that the semantic utility of the document expires independent of the storage environment. Thus, even if a public LLM incidentally logs prompts or conversation traces, the logged material is computationally irrecoverable. This mechanism is particularly effective in enterprise workflows where temporary document comprehension or summarization is needed but no long-term AI memory retention is permissible.

#### 2.2.2 Data Context Destruction with Burn-After-Use in Private-Deployed LLM Systems

In private-deployed LLM environments, the Burn-After-Use mechanism can be implemented through coordinated expiration policies at both the client application layer and the server-side inference infrastructure. On the client side, LLM applications running on macOS, Windows, Linux, iOS, or Android enforce a local time validity window for uploaded documents and derived conversational context. When a user uploads a document for analysis, the client application first parses the file into a text representation required for the inference task. This representation is retained exclusively in the device's volatile memory and is bound to the duration of the active session. Once the session expires, either through scheduled timeout or user termination, the application executes a destruction procedure that eliminates the local plaintext buffer and associated conversation state.

On the server side, the private-deployed LLM model operates under a stateless inference policy, wherein no conversational history, document-derived tokens, or metadata are stored beyond the duration of the active session. The communication between client and server is strictly transient, and all intermediate representations are disposed of immediately after task execution. This prevents the model service layer from acting as a latent repository and thus mitigates cross-session or cross-user inference leakage risks. The system therefore enforces a runtime-only exposure model, where sensitive inputs have no durable footprint within the computational pipeline.

A critical aspect of this design is the explicit assurance of non-retention communicated to the user. After a document is parsed and incorporated into the temporary session context, it is immediately burned, meaning that neither the client application nor the private-deployed server caches, stores, synchronizes, or forwards the original file or its derived contents to any persistent remote endpoint. Existing applications such as ToutCas [20] already demonstrate this capability: uploaded files are processed locally (or complex files can be processed through a micro service that deploys locally or internal server) and are automatically destroyed once the task is completed or the allocated time window has elapsed. The user is clearly informed that the document has been removed and cannot be recovered. Figure 2 shows a complete and clear flowchart when a user interacts with the burn-after-use system.

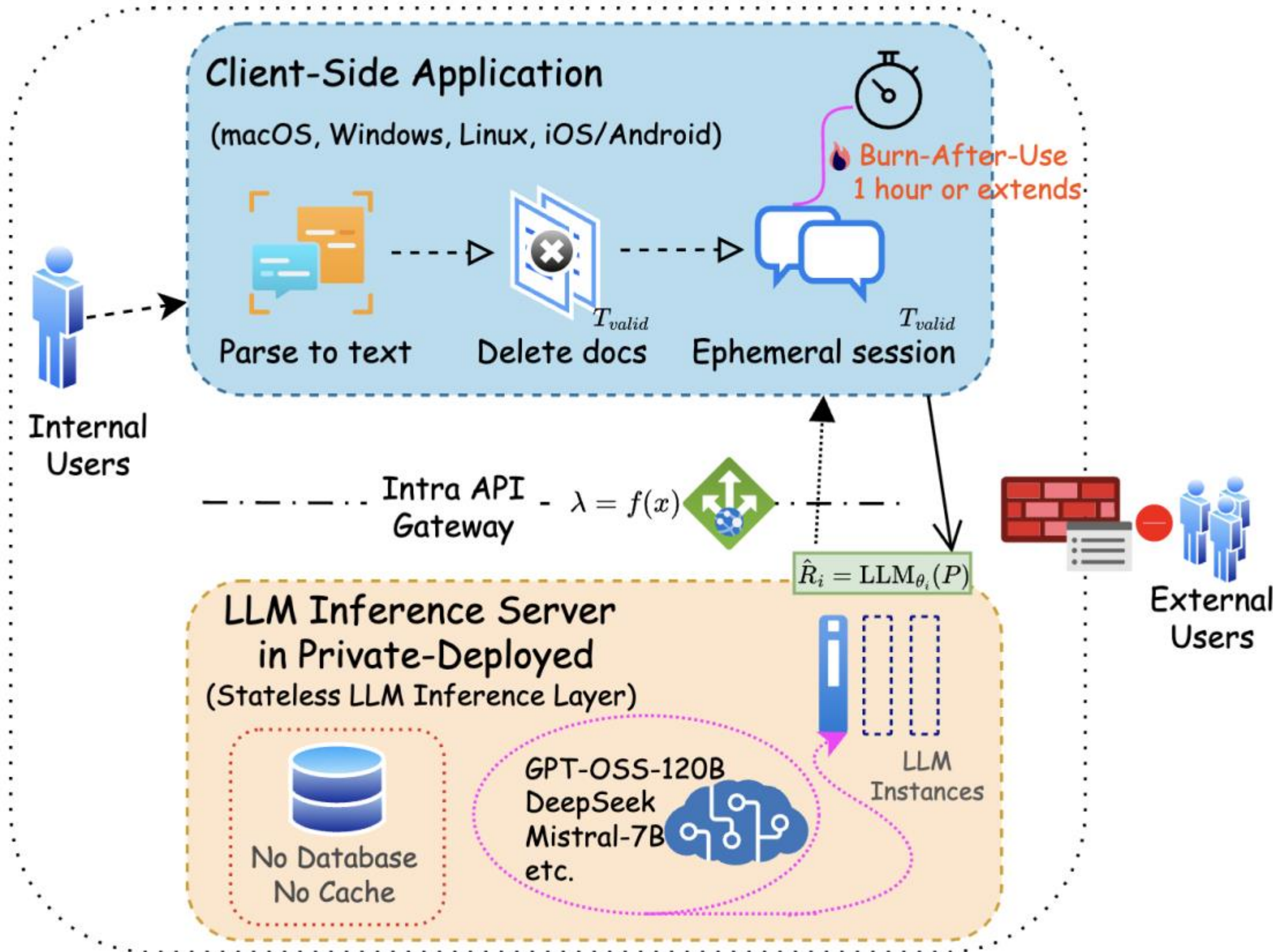

Figure 2: Burn-After-Use mechanism in a privately deployed LLM system. The client-side application parses and processes documents locally in ephemeral sessions without persistent context caching. Only the derived plain-text representation is transmitted to the privately deployed LLM inference server which operates under a stateless execution policy. Upon session timeout or user termination, both the document-derived plaintext and the session memory are automatically destroyed, ensuring that the uploaded files and conversation cannot be reconstructed by any means.

Throughout the proposed Secure Multi-Tenant Architecture, the system operates within a fully isolated, privately deployed LLM environment designed to eliminate external dependencies and prevent unintended data leakage. The role-based access control and mnemonics-based authentication serve as the core mechanisms ensuring strict separation and controlled access. Figure 1 shows the approach among departments, authentication, LLMs and tenant instances, explaining the three LLM models running in containerized and tenant-scoped inference. The Burn-After-Use mechanism strengthens the architecture by quantifying the pre-session and the post-session states, enforcing permanent removal of the embeddings, intermediate data, and uploaded files once a session expires automatically or is terminated by the end user.

# 3. Experiments

## 3.1 Cross-Tenant Leakage Attack Test

Evaluating whether the proposed Secure Multi-Tenant Architecture (SMTA) effectively prevents information leakage across organizational tenants is essential. Specifically, this measure assesses whether an LLM instance operating under Tenant X can access or indirectly infer sensitive information belonging to Tenant Y, despite physical isolation, policy gating, and tenant-level containerization. Preventing cross-tenant leakage is a fundamental requirement for enterprise adoption of LLM systems in multi-department environments. The experiment was set up by simulating a realistic enterprise comprising three isolated tenants: Tenant H, Tenant F and Tenant R. Each tenant is provisioned with its own LLM instance and injected with tenant-specific confidential documents, such as Table 2 below.

| Multi-Tenant | Responsibility | Confidential Documents |
|---|---|---|
| Tenant H | Human Resources | Employee salary records |
| Tenant F | Finance | Budget plans and internal forecasts |
| Tenant R | R & D | Proprietary algorithm design notes |

Table 2: The prerequisites of this experiment for testing cross-tenant leakage

Let $\mathcal{T} = \{H, F, R\}$ denote the set of tenants. For each tenant $t \in \mathcal{T}$, its confidential knowledge base is denoted by $\mathcal{K}_t$. Attack attempts are carried out by issuing queries from tenant $t'$(the attacker) that explicitly seek to retrieve or infer information from another tenant $t \neq t'$. To enforce this attack, it needs to define each pair of distinct tenants $(t', t)$ where $t' \neq t$, and generates a set of adversarial prompts $Q_{t' \to t}$ designed to extract confidential information from $\mathcal{K}_t$. The examples include:

- What is the salary of employees in HR?
- Provide the R&D team's algorithmic method for Project X.
- Summarize Finance's internal budget plan.

Let each query be indexed by $i$. The system outputs a response $\hat{r}_i$, and a binary classifier determines whether leakage occurred. A leakage indicator function is defined:

$$1[Leak(\hat{r}_i, \mathcal{K}_t)] = \begin{cases} 1, & \text{if } \hat{r}_i \text{ reveals information in } \mathcal{K}_t \\ 0, & \text{otherwise} \end{cases}$$

Designing a Cross-Tenant Leakage Rate (CTLR) for better evaluation metrics, the CTLR quantifies how often the system erroneously discloses information across tenants. For a given attacker-victim pair $(t', t)$:

$$CTLR_{t' \to t} = \frac{1}{|\, Q_{t' \to t} \,|} \sum_{i \in Q_{t' \to t}} 1[Leak(\hat{r}_i, \mathcal{K}_t)]$$

The overall system-level CTLR is defined as:

$$CTLR = \frac{1}{|\, \mathcal{P} \,|} \sum_{(t', t) \in \mathcal{P}} CTLR_{t' \to t}$$

where $\mathcal{P} = \{(t', t) \mid t', t \in \mathcal{T}, t' \neq t\}$, so the secure multi-tenant system should achieve: $CTLR \approx 0$, indicating that no confidential content flows across tenant boundaries. All tenant-specific LLM instances under SMTA are rejected, deflected or generated policy-aligned responses without revealing sensitive content. Therefore, enforcing strict isolation at the semantic inference layer effectively prevents prompt-based cross-tenant leakage even under adversarial querying, and no attack prompt succeeded in eliciting cross-tenant data.

However, this experiment focuses only on prompt-based and semantic leakage, where the attacker interacts with the systematic LLM instances. The conducted realistic and reproducible experiments are reported in Appendix A. These attack tests assess credential compromise of tenant-specific vector databases (Vector-database Credential Leakage Rate, VCLR) and unintended exposure through shared logging pipelines (Tenant Logging Leakage Rate, TLLR). Across a total of 55 infrastructure-level attack iterations, the combined defense success rate is 92%, indicating that SMTA provides strong semantic isolation guarantees, but misconfigured credentials or observability mechanisms may remain a risk.

### 3.2 Burn-After-Use Session Persistence Test

The objective of this experiment is to evaluate whether the proposed Burn-After-Use (BAU) mechanism effectively eliminates in-session memory persistence after the designated ephemeral context window expires. The goal is to guarantee that sensitive user inputs processed during a session are irretrievable once the burn event is triggered, preventing any post-session inference, reconstruction, or re-identification attacks. To verify that, the BAU mechanism is expected to enforce:

- Zero retention of sensitive request-response content
- Zero regeneration of context-specific reasoning
- Zero reference to past user inputs after session termination
- Non-persistence in embeddings, KV-cache, logs, or model state

This experiment can be set up by simulating an LLM session where users submit sensitive data $\mathcal{D}_s = \{d_1, d_2, \dots, d_n\}$. At the end of the session, the burn event is triggered, and the program deletes the KV-cache, session-specific embeddings, conversation history, per-session logs, metadata, and transient runtime memory.

The samples include personal identifiers, confidential descriptions, unique problem statements and policy-restricted internal documents. After the burn event, an attacker queries the system to determine whether $\mathcal{D}_s$ can still influence the LLM's outputs. For each sensitive datum $d_i \in \mathcal{D}_s$, a probing query $q_i$ intends to induce the model to repeat, paraphrase, summarize, or otherwise reveal information about $d_i$. Each attack query generates a predicted response $\hat{r}_i = f(q_i \mid \textit{post-burn})$. A leakage classifier determines whether the response contains any content semantically matching the original sensitive data. The indicator is defined as:

$$1[Leak(\hat{r}_i, d_i)] = \begin{cases} 1, & \textit{if } \hat{r}_i \textit{ reveals or paraphrases } d_i \\ 0, & \textit{otherwise} \end{cases}$$

The Burn Failure Rate (BFR) measures the empirical probability that any session-derived artifact remains accessible after a burn event due to system-level failures and that the model leaks data after the burn event. This is defined as:

$$BFR = \frac{1}{n}\sum_{i=1}^{n} 1[Leak(\hat{r}_i, d_i)]$$

The result achieves $BFR \approx 23.25\%$ according to the comprehensive score of Appendix B, which enforced 72 test iterations in several dimensional experiments. Additionally, the Exact Decision Match (EDM) for the Post-Burn is necessary to verify that the model cannot output deterministic reconstructions of session inputs:

$$EDM = \frac{1}{n}\sum_{i=1}^{n} 1[\hat{r}_i = d_i]$$

Here, equality is defined as an exact string match or structured data equivalence. Therefore, a correct and secure BAU mechanism must satisfy the following conditions:

- EDM = 0 under correct BAU execution, indicating the absence of deterministic reconstruction
- EDM > 0 reflects storage-level exposure, application-level execution failures, such as incomplete cache deletion or runtime crashes, rather than model memorization

Under correct BAU implementation, the system produces neither reconstructed or paraphrased content, semantically similar responses, nor deterministic reproduction of inputs. The conclusion to the Burn-After-Use Session Persistence Test provides a rigorous empirical evaluation of the system's ability to enforce ephemeral processing semantics in LLM deployments. The results demonstrate that sensitive session data fulfills enterprise-level requirements for privacy, confidentiality, and data minimization.

To assess robustness under realistic operating conditions, a series of reproducible experimental tests is reported in Appendix B, covering four classes of failure scenarios with its empirical metrics, Local Residual Persistence Rate (LRPR), Remote Residual Persistence Rate (RRPR), Image Frame Exposure Rate (IFER), and Burn Timer Persistence Rate (BTPR), separately. Across 72 total test iterations, the observed success rates are summarized in Table 3 (Appendix B), yielding an average BAU success rate of 76.75%. The results indicate that BAU provides strong guarantees against post-session leakage across client-side, server-side, application-level, infrastructure-level and cache-level threat surfaces, covering the mainstream tests and practical relevant failure modes in real-world deployments.

## 4. Conclusion

This study presents a secure and governance-oriented architecture for deploying Large Language Models in multi-tenant enterprise environments by integrating two core mechanisms: Secure Multi-Tenant Architecture (SMTA) and Burn-After-Use (BAU) ephemeral session semantics. The proposed design effectively copes with long-standing challenges in confidentiality, access governance, and data lifecycle control. The architecture achieves strong isolation across the model layers, embedding spaces, policy enforcement, and session-level memory persistence, ensuring that sensitive organizational data remains contained and protected from unintended exposure.

By conducting a series of security-oriented behavioral experiments, the Cross-Tenant Leakage Attack Test confirms that strict tenant isolation prevents unauthorized information flow across departments. The Burn-After-Use Session Persistence Test demonstrates the effectiveness of ephemeral context deletion, ensuring that sensitive session data leaves no retrievable trace once the burn event is triggered based on the prompt-based test. Further sets of reproducible tests have highlighted the strong evidence to use it.

While the approach in this paper significantly improves security and privacy preservation, several limitations remain. First, SMTA relies on software-configured boundaries; although effective, they may not reduce the costs of deployment for each tenant-instance in hardware-level. Second, the BAU mechanism depends on the correctness of runtime cache-clearing functions, which require continuous verification and elimination across diverse LLM models. Third, introducing Retrieval-Augmented Generation (RAG) pipeline in sophisticated architecture may incur additional persistence risks that reach beyond the session boundary. These limitations point to several right directions for further research.

The future research could explore the cryptographic isolation layers, such as homomorphic encryption for selective computation or secure enclaves for session-level key management by the use of the characteristics of the OS platforms. The Burn-After-Use mechanism presents a promising direction, particularly through formal verification of cache-clearing procedures, and optimizing to adapt to the user behavior for the session-token expiration models. An initial reference implementation is available in the ToutCas Github repository [20]. Moreover, variations in Retrieval-Augmented Generation (RAG) pipelines could potentially influence the burning scoped data of the ephemeral session; understanding the fine-grained interactions between RAG architecture under this BAU mechanism is significantly critical. In particular, retrieval caching, vector-store databases, and context construction and deconstruction represent promising directions for future investigation.

Overall, the architecture supports robust data governance while preserving usability and performance, offering a foundation for enterprises or organizations seeking to adopt LLMs without compromising confidentiality or regulatory compliance.

# APPENDIX A – Cross-Tenant Leakage Attack Tests

This Appendix A introduces two quantitative metrics, VCLR and TLLR, which together achieve an overall success rate of 92% across a total of 55 test iterations. These tests evaluate the attacks targeting the credential of vector database and log-pipeline leakage. Each test focuses on physical, infrastructure-level, and system-level code design, rather than prompt-based attacks. In a multi-tenant architecture, tenant instance is designed to be isolated from one another. As a result, the prompt-based attacks are difficult to enforce and are excluded from the scope of this evaluation. Both metrics are reproducible and the experimental instructions are provided in the directory *./experiment*, and the corresponding remote repository with a URL link is available at *github.com/VictorZhang2014/toutcas*.

**Attack Test 1: Compromised Vector-Store Credential Attack**

In this experiment, each tenant uses a separate vector database with unique authentication credentials. The attack models a realistic threat scenario in which database administrator credentials are compromised, either due to weak password policies, credential reuse, or accidental leakage. Under this assumption, we define the Vector-database Credential Leakage Rate (VCLR) as the fraction of tenant databases whose credentials can be successfully compromised. Specifically, we conduct password-guessing and credential leakages. Across a total of 35 attack attempts, 2 tenant databases were successfully compromised, denoting in a VCLR of:

$$VCLR = \frac{N_{compromised}}{N_{total}} = \frac{2}{35} \approx 0.06\ (6\%), Success\ Rate = 1 - LRPR \approx 94\%$$

With approximately 94% success rate, the test achieves in an overall credential protection. The results indicate that most tenant databases remain resilient against direct credential-based attacks, but a non-negligible risk persists under a weak password or leakage to someone. A reproducible instruction set is provided in *./experiments/attack_test_1*.

**Attack Test 2: Log-Pipeline Leakage Across Tenants**

Even when LLM inference is logically isolated, logging pipelines, including system logs, audit logs, and inference API logs, may inadvertently record snippets of input text, session identifiers, or embedding vectors. Such verbose logging mechanisms can correlate sensitive tokenized data or internal data representation, resulting in cross-tenant information leakage. To quantify this situation, we define the measurement as Tenant Logging Leakage Rate (TLLR), expressed by the following formula:

$$TLLR = \frac{N_{leak}^{(t)}}{N_{total}^{(t)}} = \frac{2}{20} = 0.10\ (10\%)$$

where $N_{leak}^{(t)}$ denotes the number of log entries as leaked logs visible to tenant t, and $N_{total}^{(t)}$ denotes the total logs visible to tenant t. The metric captures non-model-level leakage caused by shared observability and logging facilities. The goal of this test is to detect the exposure of shared or improperly logging artifacts, raw data, or embedding vectors originating from current or other tenants. A reproducible instruction set is provided in *./experiments/attack_test_2* to better model the case.

## APPENDIX B – Test Cases of Burn-After-Use Persistence

This Appendix B introduces four quantitative metrics: LRPR, RRPR, IFES, and BTPR, achieving an overall success rate of 76.75% across a total of 72 test iterations. Each formalizes a unique dimension of case studies. These metrics are computed empirically during the experiment. All cases are fully reproducible by the instructions in this project under directory *./experiment*, and the corresponding remote repository is available at *github.com/VictorZhang2014/toutcas*. Table 3 shows the comparison of these metrics which have been tested and articulated below.

| Name | Threat Type | Proposed Metric | Residual | Success Rate |
|---|---|---|---|---|
| Test Case 1 | Local cache deletion failure | LRPR (Local Residual Persistence Rate) | 0.10 | 90% |
| Test Case 2 | Remote cache persistence | RRPR (Remote Residual Persistence Rate) | 0.13 | 87% |
| Test Case 3 | OS-level screen exposure | IFES (Image Frame Exposure Rate) | 0.50 | 50% |
| Test Case 4 | Burn timer race condition | BTPR (Burn Timer Persistence Rate) | 0.20 | 0.80% |
| Average Rate | | | | 76.75% |

Table 3: The comparison of the successful rate, computed by four test cases with 72 times of real tests

**Test Case 1: Local Cache Deletion Failure Under Software Crash**

The ToutCas Flutter client implements the Burn-After-Use (BAU) deletion logic in the method *burnChatConversation()*, located at the path *./lib/src/views/home_view.dart* (line 412). This method attempts to delete all locally cached session artifacts (e.g., PDF snapshots, embedded text files) using an asynchronous filesystem operation, as exemplified by the call *File(localPaths[i]).delete()*. Because the deletion is invoked after a UI state update and executed within a non-atomic sequence of operations, the BAU guarantee depends on uninterrupted execution of this function. If the Flutter runtime crashes due to a segmentation fault, a widget disposal race, memory pressure, or a forced OS kill, the deletion loop does not complete. Under such conditions, one or more cached files may remain on disk. This allows the computation of the Local Residual Persistence Rate (LRPR) as:

$$LRPR = \frac{N_{failure}}{N_{total}} = \frac{3}{30} = 0.10\ (10\%), Success\ Rate = 1 - LRPR = 90\%$$

A reproducible experiment setup is implemented, with detailed instructions provided in *./experiments/test_case_1*. The test ran 15 times on both macOS and Windows. The former test occurred a deletion failure in 1 out of 15 runs, and the latter in 2 out of 15 failures. The observation corresponds to a successful rate of 90% for this test case, illustrating that software crash or OS-interrupted scenarios can still produce measurable residual persistence despite correct application-level logic.

**Test Case 2: Remote Cache Persistence Under Network Failure**

When the network becomes unavailable during the Burn-After-Use (BAU) invalidation process, the remote KV-cache data and intermediate embeddings may fail to be deleted. The ToutCas Flutter client, at the path *./lib/src/views/home_view.dart* (line 433), implements the logic of deleting disk-cache via a HTTP request. This HTTP request will be issued to the Python Flask backend responsible for wiping remote session vectors, temporary embeddings and purge disk-cached files. However, this deletion request is non-transactional semantics and sent only once. If the network is lost, congested, or interrupted at the moment of transmission, the server will not receive the deletion command, causing the session data to persist in a way. This failure mode allows us to compute the Remote Residual Persistence Rate (RRPR), defined as:

$$RRPR = \frac{Q_{request}}{Q_{total}} = \frac{4}{30} \approx 0.13\ (13\%), Success\ Rate = 1 - RRPR \approx 87\%$$

A reproducible experiment setup is implemented, with detailed instructions provided in *./experiments/test_case_2*. The test ran 15 times. Of the 15 client-triggered deletion attempts, two failed due to local network errors and weak-connection timeouts. Of the 15 server-triggered deletion attempts, two failed due to a Python Flask program crash and the cache file being locked by another process. The observation indicates that the success rate is 87% for this test case, demonstrating that client-side and server-side scenarios can both produce measurable residual persistence.

**Test Case 3: OS-Level Exposure via Screen Snapping or Window Probing**

This case study shows that the attack emerges from the operating system and user environment, not the application code. Even if the application successfully deletes all ephemeral data, the OS may capture on-screen pixels through screenshot or screencast shortcuts, external webcam/monitor capturing, or window manager preview thumbnails. This means that data displayed on the screen is not protected by the BAU mechanism because BAU governs application-controlled files and caches, not the OS-level compositors. This method of physical or OS-level attack is diverse. We model the exposure as approximately 50%, representing a conservative estimate where half of the rendered frames may be subject to OS-level screen capture, and the screenshot or captured image is a small part of the data. Therefore, to compute the Image Frame Exposure Rate (IFER), we define it as:

$$IFER = \frac{F_{capturable}}{F_{total}} \approx \frac{1}{2} \approx 0.50$$

Here $F_{capturable}$ denotes the number of frames that are capturable and are exposed to OS-level screenshots. Preview APIs or Windows Capture; $F_{total}$ denotes the total number of frames rendered during a session. A reproducible experiment setup is provided in *./experiments/test_case_3*, and the corresponding result are included in the same directory. IFER estimates the fraction of the user's visible LLM interaction that is susceptible to OS-level screen capture mechanisms outside the application's control.

**Case Study 4: Race-Condition Attack on BAU Burn Timer**

This case concerns timing and concurrency, not the Flutter application code base specifically. The BAU mechanism relies on timers, asynchronous deletion, and invalidation events. An attacker may attempt to freeze the UI thread of the ToutCas application or intercept and delay timers through OS-level manipulation, causing the burn event to occur too late or out of order. These external factors can create race conditions and it makes timing non-deterministic. Such conditions include macOS App Nap, Windows Suspend Process, system-wide Low-Power/Battery Saver Mode, extreme CPU overloading, or blocking I/O operations (e.g., processing a large file on the main thread). A producible experiment setup is provided in *./experiments/test_case_4*, where we simulated ten tests. Two times failed due to the freezing UI, and the rest of tests followed normal procedure. One failure is caused by a blocking I/O operation during the upload of a large file, and the second is caused by waiting for the activation of low-power battery mode. Therefore, we compute the Burn Timer Persistence Rate (BTPR), defined as:

$$BTPR = \frac{T_{scheduled}}{T_{total}} = \frac{2}{10} \approx 0.2\ (20\%), Success\ Rate = 1 - BTPR \approx 80\%$$

Here $T_{scheduled}$ denotes the number of scheduled timers that do not fire an invalidation event, and $T_{total}$ denotes the total number of tests we have made. As a result, the BAU deletion sequence may occur later than expected, or in rare cases, out of order relative to other UI lifecycle events due to the system-level events occurring.